# Multi-harmonic and special shape/pattern/template approximations of discrete signals with generally irregular arguments

*Ivan L. Andronov[1], Hanna M. Akopian[1], Vitalii V. Breus[1],*
*Lidiia L. Chinarova[1], Larysa S. Kudashkina[1], Nina V. Savchuk[1],*
*Serhii I. Iovchev[1], Vladyslava I. Marsakova[2,3], Serhii V. Kolesnikov[3],*
*Maksym Yu. Pyatnytskyy[4]*

[1] *Dep. "Mathematics, Physics and Astronomy", Odesa National Maritime University, Ukraine*
[2] *Odesa Richelieu Science lyceum, 65082, Odesa, Ukraine*
[3] *Main Astronomical Observatory, National Academy of Sciences, Kyiv, Ukraine*
[4] *Private Observatory "Osokorky", Kyiv, Ukraine*
mail: tt_ari@ukr.net,
https://www.erachair.lu.lv/conference/

> "There is no sense to elaborate new methods for the time series analysis,
> As even Ptolemy knew the Fourier Transform"
> © The anonymous referee (1996)

**Introduction**

Phenomenological modelling instead of (or complementary to) the physical one is needed because the parameters of the physical modelling are more numerous, and, typically, need additional data of other types. The simplest example is one equation between two unknowns $ab = c$, where $c$ is a "phenomenological" parameter (e.g. energy), and $a, b$ – physical parameters (e.g. power, time). So, it is impossible to determine $(a, b)$ separately, if knowing only $c$. Similarly, sometimes the physical modelling may need, say, dozens of the parameters $m$, whereas the data may be statistically optimally described by a function with much smaller $m$. The most common phenomenological models are based on the algebraical or trigonometrical polynomials for approximation of the whole data, and on the gaussian or rectangular shape of relatively short "pulses". We briefly review more complicated models with special shapes, also known as "patterns" or "templates".

The diversity of the types of deterministic and stochastic signals need adequate methods for the statistically optimal data analysis. Real detected signals are never infinite, and are discrete. Often there are large gaps between the observations, which drastically complicate power spectra, cross- (and auto-) correlation functions, functions of the parameters.

**Algorithms**

The paper, which got the referee report cited in the epigraph, was published in another (much more respectable) journal [1]. There are improved complete expressions which describe statistical properties in the complex case of "running" approximation merging separate algorithms:

- irregularly spaced discrete data
- an arbitrary covariation matrix $w_{kj}$ of the statistical errors of the measurements, which extends the "diagonal" case of the "Gaussian weights" $w_{kj}$.
- multiplicative "window" function $p(z_k, z_j)$, like in the wavelets.

Despite each of these topics are discussed separately, also with special shapes, the algorithms of the joint improvements are more complicated and have been discussed in [1,2].

The generalized version of a scalar product of the two vectors $\vec{a}$ and $\vec{b}$, which is used for further analysis, may be expressed as

$$(\vec{a} \cdot \vec{b}) = \sum_{kj=1}^{n} p(z_k, z_j) \cdot w_{kj} \cdot a_k \cdot b_j, \tag{1}$$

Here $z_j = (t_j - t_0)/\Delta t$, where $t_j$ – as a $j$ –th argument of the signal $t_j, x_j, j = 1..n$. In the "wavelet" terminology, $t_0$ is called "shift", and $\Delta t$ – "scale". Often (but not exclusively), the weight function is symmetrical $p(\pm z_k, \pm z_j) = p(\pm z_j, \pm z_k) = p(\pm z_k)$, then $t_0$ is the center of the interval of the approximation, in which the data are placed generally asymmetrically.

The test function, may be generally written as

$$\Phi(x_j; C_\alpha) = (\vec{x} - \vec{x}_C)^2 = \sum_{kj=1}^{n} p(z_k, z_j) \cdot w_{kj} \cdot (x_k - x_{Ck}) \cdot (x_j - x_{Cj}) \tag{2}$$

Here $x_{Cj}$ are "calculated" values at arguments $t_j$ according to the approximating function $x_C(t, C_\alpha)$, where $(C_\alpha, \alpha = 1..m)$ are "parameters" or "coefficients" of the mathematical model. Similarly to the basic method of the Least Squares "LS" proposed by Karl Gauss before 1805, one has to determine the set of the parameters $C_\alpha$, which minimizes the scalar function $\Phi$. This corresponds to $m$ "normal" equations $\partial \Phi / \partial C_\alpha = 0, \alpha = 1..m$. Generally, there may be a large number of solutions of these sets of the normal equations, which correspond to different values of $\Phi$. This is typical for "non-linear" basic functions, in which the coefficients are involved inside, e.g, for the mono-periodic multi-harmonic approximation of order $s$

$$x_C(t) = C_1 + \sum_{j=1}^{s}(C_{2j}\cos(2\pi j f t) + C_{2j+1}\sin(2\pi j f t)) = C_1 + \sum_{j=1}^{s} R_j \cdot \cos\left(2\pi j f(t - T_{0j})\right),$$

The frequency $f = C_{2j+2}$ is also a parameter to be determined. In the periodogram analysis, the test function $\Phi$ is computed at a grid of equally spaced values of $f$ with a recommended step $\Delta f = \Delta \varphi / sT$, where $\Delta \varphi \sim 0.05 \ll 1$, and $T$ is duration of observations. The periodogram The most popular our realization of the method is the software MCV [3,4], which also has an unique function to make a periodogram analysis taking into account a frequency-dependent trend, contrary to popular oversimplified detrending or prewhitening.

For small number of parameters $m$, the approximation(=fit) may loose some systematic components of the signal. Such a situation is called "underfit", whereas large $m$ correspond to an "overfit", where the approximation is better, following random fluctuations of the data (e.g. [5]). The statistically optimal number of parameters (of any approximation) is used practically:

- "Estetic" ="user-defined"
- ANOVA (=analysis of variances, Fischer's criterion, $p$-value, FAP=False Alarm Probability)
- Best statistical accuracy of the approximation (mean-squared or at extremum or any point)

The user-defined degree $s$ is commonly used in many computer programs, including different electronic tables. The statistically optimal values using the ANOVA – type criteria are a next step of the analysis. E,g, the catalogues of the photometrical characteristics of long-period pulsating variable stars of different subtypes were published [6,7,8]. The atlas of the phase plane $(x, x')$ curves was presented [9]. The sines and cosines may be combined into basic functions like (scaled and shifted) $\sin(\pi t)/(n\sin(\pi t/n))$

The cubic polynomial splines were used also for the periodogram analysis and a search of the period changes [11]. The splines are more local and thus sometimes may have better approximations than the trigonometrical polynomial. However, the splines depend not on the number of basic points $m$, but also on the position (shift). Thus there may be two improvements – either to find a best shift, or shift-averaged approximation. In this version, the data are still split into equal $m$ subintervals, and the function of the interval is a cubic polynomial. In reality, the argument $t$ may be split into parts, e.g. maxima and quiescence for the outbursts, or eclipses in the binary stellar systems. The simplest "spline of changing order" (1;2;1) is an "asymptotic parabola" (AP) [12] This function consists of two inclined lines ("asymptotes") connected with a parabola, so the function and its derivative is continuous. This (generally asymmetric) approximation has $m = 5$ parameters, as well as a polynomial of 4-th order. However, AP seems to have smaller systematical deviations for the maxima of majority of pulsating variable stars. For AP, there are two "non-linear" parameters $q = 2$) – the positions of the borders between the parabola and the straight line. This approximation as good for the data in a logarithmic scale (e.f. brightness in stellar magnitudes), as the slopes of the lines may be used for computation of characteristic time scales of the ascending and descending branches of the light curve. This method was used in dozens of papers of our group. The comparison between this kind of spline ($q = 2$), "symmetrical" polynomial ($q = 1$) and ordinary algebraic polynomials ($q = 0$) was presented by [13]. Generally, we use approximations with two "linear" parameters $\tilde{C}_1, \tilde{C}_2$, non-linear (shift $C_3$, scale $C_4$) and the rest describing the hape of the extremum [2]:

$$x_C[t] = \tilde{C}_1 + \tilde{C}_2 \cdot \tilde{G}\left((t - C_3), C_4, \ldots, C_{mp}\right),$$

For non-logarithmic data, the (continuous with all derivatives) asymmetrical approximation for all the hump is $\tilde{G}((t - C_3), C_4, C_5) = \frac{2}{\exp(C_4 \cdot (t - C_3)) + \exp(-C_5 \cdot (t - C_3))}$ [14]. This is an extension of the $\operatorname{sech}(z)$ popular symmetrical function. Even complicated function resembling the "log-normal" statistical distribution [15]

$$\tilde{G}((t - C_3), C_4, C_5) = \exp\left(-\ln 2 \cdot C_5 \cdot (\ln(C_4 \cdot (t - C_3) + 1))^2\right)$$

This approximation is useless for nearly-symmetrical signals because it does not converge. Another asymmetrical approximation is a second-order polynomial spline with the parameters – the borders between the intervals. For the symmetrical shape, there are symmetrical polynomials with integer and non-integer power, These and other functions (totally, 22 of 11 types) have been involved in the software MAVKA [16], available at http://uavso.org.ua/mavka . One may automatically chose the method, which corresponds to the best accuracy of the moment of the extremum.

Special shapes for the eclipses of binary systems may be split into "all phase curve", "eclipse approximation", "near-extremum parts". Obviously, the wider is the interval compared to the width, the larger $m$ is. Near extrema, the parabolic fit may be enough, with an increasing $m$ for wider intervals. For the "almost" flat minima correponding to transits of exo-planets, or total eclipses, there are three "wall-supported" functions [17]. The most common function for the spectral lines is a gaussian $\exp(-z^2/2)$ was also used by [18] with some modifications like $\exp(1 - \cosh(z))$. These shapes have infinite width, so it is impossible to determine borders of the minimum which are requested in the official catalogs of variable stars, mainly due to absence of adequate software. Thus, in 2010, we proposed the "New Algol Variable" (NAV) function [19], which is also effective for other types of the eclipsing variables (EA, EW) with smooth curves [20]. There may be minor improvements [18, 21], which increase $m$ often without making significantly better fit. These studies are partially made within the "Inter-Longitude Astronomy" (ILA) project.